\documentclass[sigconf]{acmart}
\usepackage{graphicx} 

	\author{Julien Monteil}

	\affiliation{%
		\institution{Amazon Machine Learning}
		\city{Brisbane}
		\country{Australia}	
  }
	
	\author{Volodymyr Vaskovych}
	\affiliation{%
		\institution{Amazon Machine Learning}
  		\city{Melbourne}
		\country{Australia}}
	
	\author{Wentao Lu}
	\affiliation{%
		\institution{Amazon Machine Learning}
		\city{Sydney}
		\country{Australia}
  }
  	\author{Anirban Majumder}
	\affiliation{%
		\institution{Amazon Machine Learning}
		\city{Bangalore}
		\country{India}
  }
	
	\author{Anton van den Hengel}
	\affiliation{%
		\institution{University of Adelaide}
		\city{Adelaide}
		\country{Australia}}
	
\AtBeginDocument{%
	\providecommand\BibTeX{{%
			\normalfont B\kern-0.5em{\scshape i\kern-0.25em b}\kern-0.8em\TeX}}}

\usepackage{siunitx}
\usepackage{amsmath}
\usepackage{graphicx}
\usepackage[utf8]{inputenc} 
\usepackage[T1]{fontenc}    
\usepackage{hyperref}       
\usepackage{url}            
\usepackage{booktabs}       
\usepackage{amsfonts}       
\usepackage{nicefrac}       
\usepackage{microtype}      
\usepackage{xcolor}  

\newcommand{\manet}{{\textsc{MARec}}} 
\newcommand{\cut}[1]{}

\begin{document}

	\title{MARec: Metadata Alignment for cold-start Recommendation}

	\begin{abstract}
		For many recommender systems, the primary data source is a historical record of user clicks. The associated click matrix is often very sparse, as the number of users $\times$ products can be far larger than the number of clicks. Such sparsity is accentuated in cold-start settings, which makes the efficient use of metadata information of paramount importance. In this work, we propose a simple approach to address cold-start recommendations by leveraging content metadata, Metadata Alignment for cold-start Recommendation (\manet). We show that this approach can readily augment existing matrix factorization and autoencoder approaches, enabling a smooth transition to top performing algorithms in warmer set-ups. Our experimental results indicate three separate contributions: first, we show that our proposed framework largely beats SOTA results on 4 cold-start datasets with different sparsity and scale characteristics, with gains ranging from +8.4\% to +53.8\% on reported ranking metrics; second, we provide an ablation study on the utility of semantic features, and proves the additional gain obtained by leveraging such features ranges between +46.8\% and +105.5\%; and third, our approach is by construction highly competitive in warm set-ups, and we propose a closed-form solution outperformed by SOTA results by only 0.8\% on average.
		
	\end{abstract}

	\maketitle

\section{Introduction}
	
Recommender systems have been a hot topic of interest in the research community, even more so since we interact with large-scale recommender systems in our daily lives. Significant effort has been dedicated to surpass the performance of state-of-the-art methods on public datasets, focusing on learning the best representation of the sparse user-item interaction matrix. Less interest has been vested on solving the different yet practical problem of cold-start recommendation which consists of providing relevant recommendations for new users and items as soon as they appear on the platform. 

Cold-start is a significant challenge for recommender systems as traditional collaborative filtering techniques rely heavily on past user interactions which are not available for new users or items. As a result, existing systems that are optimized for warm (previously seen) users may generalize poorly, if at all, to the cold-start scenario. As recommender systems constantly experience flurry of new items and users, high quality cold-start recommendations are essential for a great user experience and long-term engagement.
It is important to highlight that the cold-start scenario extends beyond just looking at the ranking performance on new items (or users); relevant recommendations are required for the mixed scenarios as well i.e. with a blend of warm and cold items (users) with varying amount of historical data.
	

Amongst early works on cold-start recommendation modeling figure the works of~\cite{gantner2010learning,ning2012sparse} which respectively map metadata features to Matrix factorization latent factors, and leverage item metadata reconstruction for regularization. More recent techniques for cold-start modeling include meta learning to combine user history with item representation~\cite{vartak2017meta}, Variational AutoEncoders~\cite{xu2022alleviating}, and the novel idea of content-aware hashing~\cite{zhang2018discrete}. However, while current approaches for cold-start modeling demonstrate good performance in cold-start scenarios, their effectiveness tends to decline significantly when applied to warm-start settings. In an industrial setup, managing separate models for cold-start and warm-start proves cumbersome. This not only adds to the overhead of maintaining multiple models in production but also poses a challenge to ensure a consistent and high-quality pool of recommendations for users when they transition between the cold and warm regime. In this paper, we present \manet\ ({\bf M}etadata {\bf A}lignment for cold-start {\bf Rec}ommendation), a new approach that excels beyond current methods in cold-start recommendation all the while being competitive with state-of-the-art (SOTA) techniques for warm-start recommendation.

 
Our approach builds upon the existing body of literature in particular the hybrid approaches of merging content-based and collaborative filtering techniques~\cite{koren2008factorization,lam2008addressing} as well as the recent works on retrieval-augmented systems~\cite{borgeaud2022improving}. Our core contribution is a novel yet simple idea of aligning item-item similarities
	estimated from user clicks and from the item metadata features via a regularization term that pushes the cold-start items onto the same similarity space. In this way, we leverage SOTA models for warm-start recommendation to learn an alignment of the feature-based similarities to the user-item interaction similarities, which provides a high-performing cold-start recommender. 
 
 \subsection{Contributions}
 
To the best of our knowledge, the alignment framework we propose differs from the existing body of literature, as it can augment any top performing Matrix Factorization and (variational) autoencoders approaches, it makes the most of any additional metadata features we can leverage across modalities (image, categorical, textual), and yet it achieves impressive results despite its simplicity. We present the following contributions:
	%
	\begin{enumerate}
\item We present a novel algorithm (\manet) for cold-start recommendation that achieves state-of-the-art performance on several benchmarking datasets. Our algorithm combines embeddings learned from item and customer metadata with the user-item click matrix and gives closed-form update equations.
\item We show that \manet\ beats SOTA techniques on four cold-start benchmarking datasets with different sparsity and scale characteristics, with gain ranging from +8.4\% to +53.8\% on standard ranking metrics.
\manet\ achieves this performance while being orders of magnitude faster, in terms of training time, compared to the best performing baseline.
\item We provide a study on the utility of Large Language Models (LLMs) embeddings and demonstrate that the additional gain obtained by leveraging semantic features ranges between +46.8\% and +105.5\%.
\item Finally, \manet\ enables a transition to near-SOTA performance in warm set-ups, and we introduce a closed-form solution outperformed by SOTA results on warm datasets by only 0.8\% on an average.

	\end{enumerate}
	%
	%
	%
\section{Related Works}
The historical  way to perform collaborative filtering with click datasets is via neighbourhood methods which consists of computing similarities across items or users, where the similarity is evaluated as a dot product based on click behaviour, and possibly item and user metadata~\cite{4781121}. The Netflix competition and subsequent research works have shown that linear and matrix factorization techniques tend to outperform neighbourhood methods~\cite{koren2009matrix,ning2011slim,steck2020admm}. A push for neural approximations of matrix factorization took place~\cite{he2017neural,covington2016deep}, with the conclusion that the proposed architectures fail to learn a better nonlinear variant of the dot product and are outperformed by careful implementations of matrix factorization~\cite{dacrema2019we,rendle2020neural,rendle2022revisiting}. More recently, the variational autoencoder approach~\cite{liang2018variational} was proposed, but the least squares approximation of SLIM~\cite{ning2011slim}, which is a linear autoencoder with projections to ensure the zero diagonal constraint (EASE)~\cite{steck2019embarrassingly,steck2020admm}, still beat it on 2 out 3 public datasets. It was shown in~\cite{kim2019enhancing} that augmenting the VAE approach with flexible priors and gating mechanisms led to SOTA performance. The VAE approach was also refined and ensembled with a neural EASE to achieve SOTA~\cite{vanvcura2021deep} on MovieLens20M and Netflix. The success of Deep Cross Networks on learning to rank tasks~\cite{qin2021neural} also showed the importance of explicitly encoding the dot product in the network architecture.

Comparatively, less research has been done on the usefulness of the SOTA collaborative filtering algorithms for cold-start recommendations. An early effort to incorporate item metadata into the SLIM family of models was the one of~\cite{ning2012sparse}, where it is shown that adding a regularization term pushing towards the reconstruction of the feature information matrix, could help achieve gain on some datasets. Early methods addressed this fusion problem by proposing hybrid models merging neighborhood and collaborative filtering methods~\cite{koren2008factorization,lam2008addressing}, where the core idea relies on weighting a content-based filtering term and a collaborative filtering term in the objective function, thereby training the model to leverage content representations when preferences are not available. In~\cite{popescul2013probabilistic}, the sparsity problem is mitigated for cold start by developing a model at the word level, not at the document level. More recently, it was shown in~\cite{chen2019differentiating} that utilising different regularization weights to the latent factors associated to users and items shows consistent benefits for addressing cold-start in matrix factorization. In~\cite{volkovs2017dropoutnet}, a dropout technique is leveraged to set some click inputs to zero in order to force the model to reconstruct the relevance scores without
seeing the warm inputs. A meta learning algorithm deployed in a production system was proposed in~\cite{vartak2017meta} to combine user history representation with item representation. 

Another recent work~\cite{CLCRec} relies on maximizing the mutual information between feature representations and user-item collaborative embeddings, and it is shown that the presented technique beats a number of previous approaches~\cite{du2020learn,7410843} building on the same intuition. This approach was extended in~\cite{equal}, which addresses the item feature shift generated by cold-start items via an equivariant learning framework. The variational framework of~\cite{liang2018variational} is adapted in~\cite{xu2022alleviating} for addressing cold-start in particular to obtain reasonable priors so that to IDs with the similar attributes can naturally cluster together in the latent embedding space. In~\cite{zhang2018discrete}, a content-aware hashing approach named Discrete Deep Learning (DDL), which first relies on building an user-item preference model based on hash codes, is proposed to alleviate data sparsity and cold-start item problem. A hybrid model is proposed in~\cite{bernardis2021nfc} where a neural model is trained to learn feature embeddings and item similarities from the similarities pre-learned from a collaborative filtering model, e.g. SLIM. In~\cite{zhao2022improving}, a method is proposed to warm-up cold item embeddings by modelling the relationship between item id and side information in the latent space. 

The ensembling of complementary models remains another possibility for boosting cold-start recommendations, e.g., the winning solution of the 2018 ACM challenge on a cold-start music recommendation dataset was the score ensembling of a content-aware deep cross network and a gradient boosted decision tree~\cite{bai2017incorporating}. Finally, retrieval-augmented models present similar architectures to hybrid recommender systems, e.g. in ~\cite{khandelwal2019generalization} where the linear interpolation of a pre-trained LLM with a k-nearest neighbors model led to SOTA results on Wikitext-103, in~\cite{borgeaud2022improving} where they use a chunked cross-attention mechanism to condition the training of the language model to encoded neighbours, and in~\cite{long2022retrieval} where a retrieval module to return the top-k text descriptions associated to the image helps achieves SOTA results for visual long tail classification.

In the context of the above literature, our paper situates itself in the category of hybrid recommender systems, at the intersection of collaborative and content-based filtering. Differently to the existing literature on hybrid recommender systems, we put a specific focus on achieving scalability, and on leveraging the benefits of collaborative filtering techniques, as we transition to warmer situations. Typical hybrid approaches tend to complicate the learning by forcing the balancing of multiple objective terms, in addition to training content representations. In particular, the content part of the objective function is trained to explain the observed data, hence it may model content aspects that end up hurting cold-start recommendations. In our framework, we provide a flexible scalable approach that learns to only leverage the relevant features of content metadata, and aligns the notion of item similarities independently of whether it comes from click data or item metadata. 

 \section{Modelling Approach}
	Let $\mathcal{U}$ and $\mathcal{I}$ denote the set of users and items respectively.  We assume that the data is available as a sparse matrix $\mathbf{X}\in\mathbb{R}^{|\mathcal{U}|\times|\mathcal{I}|}$ where $X_{ij}$ denotes the interaction (click, rating) between user $i \in \mathcal{U}$ and item $j \in \mathcal{I}$. A value of $1$ in $X_{ij}$ indicates that the user interacted with the item while a value of $0$ implies that no interaction has been
	observed. We use the shorthand notation $\mathbf{X_{i.}}$ and $\mathbf{X_{.j}}$ to denote the $i^{\text{th}}$ row 
	and $j^{\text{th}}$ column 
	of $\mathbf{X}$. 
	Similarly, we use $x_i$ to denote the $i^{\text{th}}$ element of a vector $x$. 
	
	For each item, we have access to a list of $N$ attributes representing metadata associated with each item such as item name, description, brand, etc. We assume that the $k^{\text{th}}$ attribute across all items has a compact feature representation $\mathbf{F^{(k)}}\in\mathbb{R}^{|\mathcal{I}|\times n_k}$ where $n_k$ is the dimension of the feature space. Finally, $\mathbf{F}\in\mathbb{R}^{|\mathcal{I}|\times K}$ represents the feature vector of items with $K=\sum_{k=1}^N n_k$. 
	
	\subsection{\manet}
	\manet\ consists of three components: 1) a backbone model $f^{\text{B}}$ which is a latent network that learns a
	low-dimensional representation of the
	interaction data $\mathbf{X}$ and its reconstruction, 2)  an embedding model $f^{\text{E}}$ that encodes a dense representation of
	the item metadata and 3) an alignment model $f^{\text{A}}$ that aligns the metadata representation with the click history. This aligned representation can then be fused with the backbone model to reconstruct the click history. The backbone model $f^{\text{B}}:\mathbb{R}^{|\mathcal{U}|\times|\mathcal{I}|} \rightarrow\mathbb{R}^{|\mathcal{U}|\times|\mathcal{I}|}$ can be chosen among the well-performing collaborative filtering algorithms of matrix factorization and autoencoder families, e.g., ~\cite{kim2019enhancing,steck2019embarrassingly}. Below we describe each component in detail. 
	
\begin{figure}[h!]
  \centering
  \includegraphics[height=7.0cm]{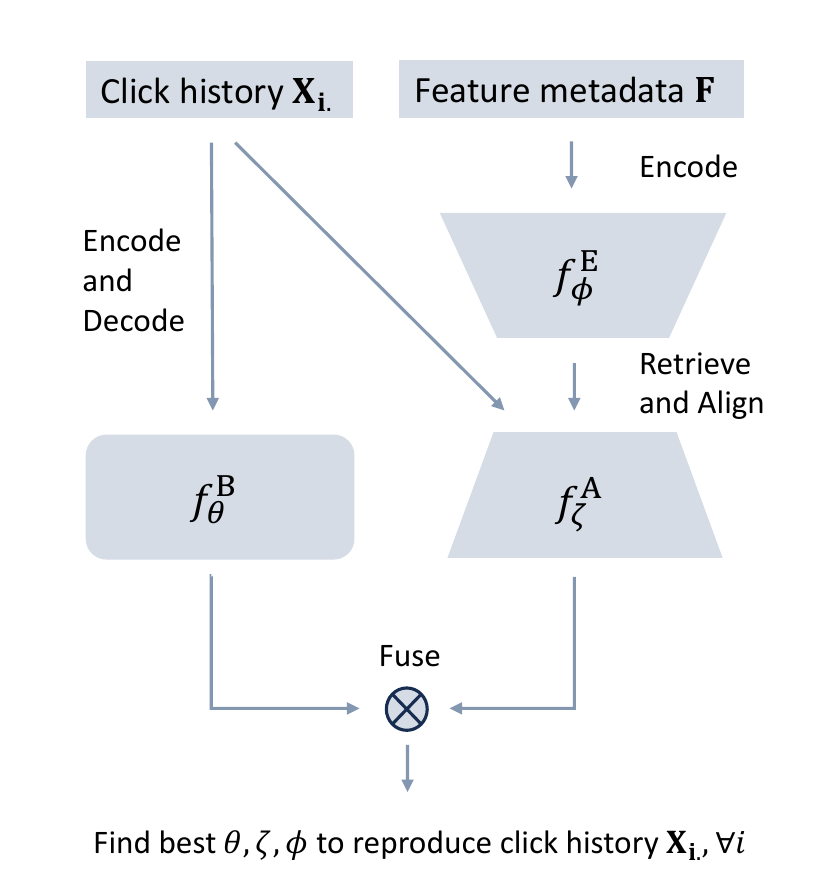}
  \caption{Simplified architecture of \manet. The embedding model $f^{\text{E}}_{\phi}$ embeds each item metadata in a joint embedding space, and the alignment function $f^{\text{A}}_{\zeta}$ retrieves and align the similar items to the similarity space provided by the backbone model $f^{\text{B}}_{\theta}$. The architecture is trained with a reconstruction loss on the click history.}
  \label{fig:diagram}
\end{figure}
	
	 The embedding model $f^{\text{E}}:\mathbb{R}^{|F|\times|\mathcal{I}|}\rightarrow\mathbb{R}^{|E|\times|\mathcal{I}|}$ learns a dense representation of item metadata so that items with similar
	 interaction pattern are mapped closer in the embedding space. The alignment model $f^{\text{A}}:\mathbb{R}^{|\mathcal{U}|\times|\mathcal{I}|}\times\mathbb{R}^{|E|\times|\mathcal{I}|}\rightarrow\mathbb{R}^{|\mathcal{U}|\times|\mathcal{I}|}$ is responsible for aligning interaction data with the item-item similarity estimated via the embedding model $f^{\text{E}}$. Figure~\ref{fig:diagram} presents the high-level overview of \manet.  
	 
	 The model is learned by minimizing a weighted combination of the backbone model loss and the alignment loss:
	\begin{equation}
	    \min_{\theta,\phi,\zeta}~ L_{\text{C}}\left(\mathbf{X},f_{\theta}^{\text{B}}(\mathbf{X})\right)+L_{\text{D}}\left(f_{\theta}^{\text{B}}(\mathbf{X}),f_{\zeta}^{\text{A}}\left(\mathbf{X},f_{\phi}^{\text{E}}(\mathbf{F})\right)\right)
	    \label{global}
	\end{equation}
	where $\theta, \phi, \zeta$ are the model parameters of individual components of \manet. Note that $L_{\text{C}}$, $L_{\text{D}}$, can be any loss function, e.g., quadratic, cross entropy, cosine, multinomial likelihood, possibly augmented with regularization terms and constraints, as in the case of autoencoders to prevent overfitting~\cite{steck2020autoencoders}. 
	While the first term of the loss function is the familiar one leveraged in matrix factorization and autoencoder techniques, the second term can be understood as a regularization to push the collaborative filtering similarities towards the learned metadata embeddings. This is particularly beneficial for cold-start items as they are under-represented in the interaction data. It is to be noted that \manet\  modifies only the training objective by adding a regularization term. Once trained, the backbone model can be used for inference as usual. \manet\ can be seen as a generalization of the merging of collaborative and content-based filtering techniques and admits a closed-form solution with appropriate choice of the components. 
	
	
	
	\subsection{Embedding Model} 
	The embedding model can be either the identity function $f^E=\mathbf{I}$, keeping the encoded metadata features $\mathbf{F}$ as they are, with $|E|=|F|$, or a Siamese network~\cite{10.5555/2987189.2987282} trained on the processed click data $\mathbf{X}$ which aims at learning joint metadata embeddings.
	We follow a similar process to~\cite{mcauley2015image} to train the Siamese network. For each item $I_j\in\mathcal{I}$, we select $k$ items with the highest cosine similarity $\mathbf{X_{.j}}^{T}\mathbf{X}$ and assign them a label of one. Similarly, we sample $k$ items with the lowest cosine similarities and label
	them as zero. This gives us a dataset of similar and dissimilar pairs of the form 
	$\left(\mathbf{F_{.j}},S_j^+,S_j^-\right)$ where $S_j^+$ is the set of
	positive pairs and $S_j^-$ is the set of negative pairs. We then train the network by minimizing the contrastive loss~\cite{10.5555/3524938.3525859}:
		\begin{equation}
	   \sum_j \sum_{k^+ \in S_j^+} \left[- \log \frac{\exp{(z_{k^+}^T \cdot z_j/\tau)}}{\exp{(z_{k^+}^T \cdot z_{j}^T/\tau)
	    + \sum_{k^- \in S_j^-} \exp{(z_{k^-}^T \cdot z_{j}^T/\tau)}}} \right]
	\end{equation}
where $z_j = f^{\text{E}}(\mathbf{F_{.j}}), z_{j^+} = f^{\text{E}}(\mathbf{F_{.j^+}}), z_{j^-} = f^{\text{E}}(\mathbf{F_{.j^-}})$ are the latent representations learned by
the embedding model and $\tau > 0$ is a temperature hyper-parameter.
The embedding layer presents the benefit of directly optimizing for the (dis)similarities of items which in turn is leveraged by the alignment model.
	
	\subsection{Alignment Model}
\cut{
	First, we go from the embedding space to similarity values simply by applying the smoothed cosine transform $g:\mathbb{R}^{|\mathcal{I}|\times|E|}\rightarrow\mathbb{R}^{|\mathcal{I}|\times|\mathcal{I}|}$, see e.g.~\cite{dacrema2019we}:
	\begin{equation}
	g\left(f^{\text{E}}(\mathbf{F})\right)=\mathbf{G},\quad G_{ij}=\frac{f^{\text{E}}(\mathbf{F_{.i}})f^{\text{E}}(\mathbf{F_{.j}})}{\|f^{\text{E}}(\mathbf{F_{.i}})\|\|f^{\text{E}}(\mathbf{F_{.j}})\|+\delta},\quad i,j\in\mathcal{I}\label{cosine}
	\end{equation}
	where $\delta$ is a smoothing term used to lower the similarity between items having only few interactions~\cite{dacrema2019we}. We use the same cosine transform in the Siamese network before computing the softmax with temperature scaling. By applying this transform for all items $i,j\in\mathcal{I}$ we are computing a similarity transform. 
 }
The alignment layer is a critical component of \manet\ with the objective of
aligning the observed click matrix $\mathbf{X}$ with an item-item similarity measure
defined over the metadata embeddings encoded by the embedding layer $f^{\text{E}}$. The item-item similarity is typically captured via a smoothed cosine transform~\cite{dacrema2019we} $g:\mathbb{R}^{|\mathcal{I}|\times|E|}\rightarrow\mathbb{R}^{|\mathcal{I}|\times|\mathcal{I}|}$,
	\begin{equation}
	g\left(f^{\text{E}}(\mathbf{F})\right)=\mathbf{G},\quad G_{ij}=\frac{z_i^T\cdot z_j}{\|z_i\| \cdot\|z_j\|+\delta},\quad i,j\in\mathcal{I}\label{cosine}
	\end{equation}
where $\delta > 0$ is a smoothing term used to lower the similarity between items with very sparse feature vectors~\cite{dacrema2019we}. When the embedding model is the identity function, the smoothed cosine transform simply computes similarities by taking the complete feature vector $\mathbf{F}$ as input, and this is known as itemKNN~\cite{10.1145/963770.963776}. However, it is expected that some attributes in the metadata are more predictive of item similarity than others. Hence an alternative is to compute embedding of each attribute $\mathbf{F}^{(\text{k})}$ separately, and then weigh attribute representations at a later stage. More specifically, starting from the encoded representation of the $\text{k}^{\text{th}}$ attributes, $z_j^{(k)}=f^{\text{E}}(F_{.j}^{(k)})$, we define item-item similarity
based on the $\text{k}^{\text{th}}$ attribute as
\begin{equation}
    G_{ij}^{(k)}=\frac{z_i^{(k)}\cdot z_j^{(k)}}{\|z_i^{(k)}\|\cdot \|z_j^{(k)}\|+\delta},\quad i,j\in\mathcal{I},
\end{equation}
which leads to $N$ similarity matrices mixed via a quadratic form:
\begin{equation}
    	g\left(f^{\text{E}}(\mathbf{F})\right)=\sum_{k<l} \left(\mu_k\cdot \mathbf{G^{(k)}}+\mu_{kl}\cdot \mathbf{G^{(k)}}\mathbf{G^{(l)}} \right),
\label{weighted_cosine}
\end{equation}
where the scalar coefficients $\mu_k\geq 0$ and $\mu_{kl}\geq 0$ are model parameters that learn importance of first and second order similarities across the feature sub-spaces. This quadratic form explicitly model non-linear similarities across feature vectors, e.g. between actors, directors, genres, in the case of movie recommendation. Finally, we define the alignment function as,
			\begin{equation}
		    f^{\text{A}}\left(\mathbf{X},f^{\text{E}}(\mathbf{F})\right)=\alpha \cdot \mathbf{X}  \ g\left(f^{\text{E}}(\mathbf{F})\right)\mathbf{D^{R}}\label{ali1}
		\end{equation}
		where $\alpha\in\mathbb{R^*_+}$ is a scaling coefficient, $\mathbf{D^R}=[d_j]_{j \in [|\mathcal{I}|]}\in\mathbb{R}_{+}^{|\mathcal{I}|\times|\mathcal{I}|}$ is a regularizing diagonal matrix that weights the importance of the metadata embeddings as a function of the number of clicks per item. 
		The entries in $\mathbf{D^R}$ are set as 
		$d_j = h\left(\sum_j \mathbf{X_{.j}}\right)$, with $h:\mathbb{N}\rightarrow\mathbb{R}^{+}$, a non-negative monotone decreasing linear or exponential function. Note that a monotone decreasing function is used to mitigate the popularity bias problem~\cite{10.1145/3437963.3441820}, thereby giving more importance to cold-start items during model training. The significance of popular items is nevertheless captured in the backbone model that we describe next.

\subsection{Backbone Model}
	
	Our framework supports any model of the autoencoder and matrix factorization families as we aim to align the item embeddings with the learned similarities from the click matrix itself. We consider 3 competitive baselines in this work, the multi-VAE model of~\cite{liang2018variational}, a collective version of the EASE model~\cite{steck2019embarrassingly,ning2012sparse}, and a modified SLIM model that we introduce. 
	
\paragraph{Variational autoencoders.} For each user $i\in\mathcal{U}$, the model samples a K-dimensional latent representation $\xi_i$ from a standard Gaussian prior. The latent representation $\xi_i$ is then passed to a multilayer perceptron $f_\theta(\xi_i)$ to produce a probability distribution over the set of items $\pi(\xi_i)$ from which the click history $\mathbf{X_{i.}}$. For each user $i$, we then adapt the multinomial log-likelihood to be maximized and the global objective of equation~\eqref{global} can be modified to: 
\begin{equation}
\begin{split}
\min_{\theta}-\log p_{\theta}(x_i|\xi_i) &= \sum_{j\in\mathcal{I}} -X_{ij}\log(\pi_j(\xi_i))-b_{ij}\log(\pi_j(\xi_i)),\\
\mathbf{b_i}&=f^{\text{A}}\left(\mathbf{X_{i.}},f^{\text{E}}(\mathbf{F})\right).
\end{split}
\label{genericVAE}
\end{equation}
To learn the parameters $\theta$ of the generative model, we need to approximate the intractable posterior distribution $p(\xi_i|X_{i.})$ and may use variational inference as in~\cite{liang2018variational,kim2019enhancing,xu2022alleviating}. We note that the second term added to the objective takes care of both the alignment and fusion by adapting the learned probability distribution over items to the metadata-based estimates. Here, the multi-VAE model of~\cite{liang2018variational} was considered but we can readily consider its recent enhancements, H+Vamp~\cite{kim2019enhancing} and VASP~\cite{vanvcura2021deep}, which now occupy the first 2 slots of the MovieLens20M and Netflix leaderboards.

\paragraph{EASE} We adapt the EASE model~\cite{steck2019embarrassingly} to our global objective of equation~\eqref{global} which becomes:

	\begin{equation}
	\begin{split}
	\min_{\Theta} &\| \mathbf{X}-\mathbf{X}\mathbf{\Theta}\|_F^2+ \lambda_1\|\mathbf{\Theta}\|_F^2+\|\mathbf{X}\mathbf{\Theta}\mathbf{D^R}-f^{\text{A}}\left(\mathbf{X},f^{\text{E}}(\mathbf{F})\right)\|_F^2\\
	&\text{s.t.}\quad \text{diag}(\mathbf{\Theta})=0,
	\end{split}
	\label{genericease}
	\end{equation}
	
	with $\mathbf{\Theta}\in\mathbb{R}^{|\mathcal{I}|\times|\mathcal{I}|}$ and $\beta\in\mathbb{R}$ a scalar coefficient. Note that we may add the collective term $\lambda_0\|\mathbf{F}-\mathbf{F}\mathbf{\Theta}\|_F^2$ to the objective function as in~\cite{ning2012sparse}. This optimization policy admits a closed formed solution via the use of a Lagrangian multiplier heuristics that sets $\text{diag}(\mathbf{\Theta})$ to zero~\cite{steck2019embarrassingly} and the solution is written:
	\begin{equation}
	\begin{split}
	\mathbf{P} &= \left(\mathbf{X}^{\text{T}}\mathbf{X}+\lambda_0\mathbf{F}^{\text{T}}\mathbf{F}+\lambda_1 \mathbf{I}+\mathbf{X}^{\text{T}}f^{\text{A}}\left(\mathbf{X},f^{\text{E}}(\mathbf{F})\right)\right)^{-1},\\
	\tilde{\mathbf{\Theta}} &= \mathbf{P}\left(\mathbf{X}^{\text{T}}\mathbf{X}+\lambda_0\mathbf{F}^{\text{T}}\mathbf{F}+\mathbf{X}^{\text{T}}f^{\text{A}}\left(\mathbf{X},f^{\text{E}}(\mathbf{F})\right)\right),\\
	\mathbf{\Theta} &= \tilde{\mathbf{\Theta}}-\mathbf{P}\frac{\text{diag}(\tilde{\mathbf{\Theta}})}{\text{diag}(\mathbf{P})}.
	\end{split}
	\label{genericEASE}
	\end{equation}

\paragraph{Modified SLIM} 
We do observe that the least squares framework of~\cite{steck2019embarrassingly} could be further improved for two reasons: first, by giving less weight to negative samples, as we observe that matrix factorization results in~\cite{rendle2020neural} were boosted by a simple negative subsampling strategy; second, by relaxing the projection heuristics used to enforce the $\text{diag}(\mathbf{\Theta})=0$ constraint. We can then write a constraint-free optimisation policy and the backbone model loss of equation~\eqref{global} becomes: 

\begin{equation}
\min_{\mathbf{\Theta}} \sum_{i} \|\mathbf{W_i}^{\frac{1}{2}} (\mathbf{X_{.i}}-\mathbf{X}\mathbf{\Theta_{.i}})\|_F^2+ \lambda_1\|\mathbf{\Theta_{.i}}\|_F^2 +\gamma_1|\Theta_{ii}|^2,
\label{wls}
\end{equation}
where $\mathbf{W_i}\in\mathbb{R}^{|\mathcal{U}|\times|\mathcal{U}|}$, $i\in\mathcal{I} $, are the diagonal matrices of the weights $w_0$ and $w_1$ for positive and negative samples, and $\gamma_1\sum_i|\Theta_{ii}|^2$ is the new regularization term that relaxes the constraint $\text{diag}(\mathbf{\Theta})=0$ but serves the same goal of penalising the trivial $\mathbf{\Theta}=\mathbf{I}$ solution. The closed-form solution to the global objective of equation~\eqref{global} is now written, $\forall i\in\mathcal{I}$:

\begin{equation}
\begin{split}
\mathbf{\Theta_{.i}} =& \left(\mathbf{X}^{\text{T}}\mathbf{W_{i}}\mathbf{X}+\mathbf{X}^{\text{T}}\mathbf{W_{i}}\mathbf{B}+
\lambda_1 \mathbf{I}+\mathbf{\Gamma^{(i)}}\right)^{-1}\left(\mathbf{X}^{\text{T}}\mathbf{W_{i}}\mathbf{X_{.i}}+\mathbf{X}^{\text{T}}\mathbf{W_{i}}\mathbf{B_{.i}}\right),\\
\mathbf{B}=& f^{\text{A}}\left(\mathbf{X},f^{\text{E}}(\mathbf{F})\right),
\end{split}
\label{genericMSLIM}
\end{equation}
with $\mathbf{\Gamma^{(i)}}$ the diagonal matrix made of zeros except at position $i$ where $\Gamma^{(i)}_{ii}=\gamma_1$, and $\mathbf{B_{.i}}$ the $i^{\text{th}}$ column of the alignment matrix.

	\section{Experiments}
	
	We leverage cold and warm dataset splits used in previous works where SOTA results were reported, for reproducibility reasons. We plan to release the \manet\ code base upon acceptance of our work.
	
	\subsection{Datasets} 

\paragraph{Cold splits} Experimental evaluation of \manet\ requires datasets with varying sparsity level, and where item metadata can be leveraged for cold-start recommendations. We select popular dataset splits explored in the cold-start recommendation literature~\cite{bernardis2021nfc,CLCRec}, as shown in Table~\ref{cdatasets}.

		\begin {table}[h!]
	\begin{center}
		\resizebox{\columnwidth}{!}{
			\begin{tabular}{c| c c c c}
				&\# users &\# items & \# item features & Interactions (\%)\\ [0.5ex] 
				\hline\hline
				Amazon Video Games~\cite{bernardis2021nfc}&33858 & 12463& 5&0.06\\ 
								\hline
        				Netflix~\cite{netflix}~\cite{bernardis2021nfc}&469986&9503&6&1.19\\
\hline
				MovieLens10M~\cite{ml10m}&55485&5986&5&1.34\\
				\hline
				MovieLens Hetrec~\cite{bernardis2021nfc}&2107&6234&6&3.10\\
    \hline

		\end{tabular}}
	\end{center}\caption{Characteristics of cold-start data splits used in our experiments ordered by sparsity level.}\label{cdatasets}
	\end{table}
	
The dataset splits were generated from binarized click interaction data to reproduce a cold-start item scenario as per the procedure described in the initial work of~\cite{sharma2015feature}. For the Amazon Video Games and MovieLens Hetrec datasets, we leverage the publicly available splits\footnote{https://github.com/cesarebernardis/NeuralFeatureCombiner} of~\cite{bernardis2021nfc}, for which 60\% of items are kept for training, 20\% for validation and 20\% for test. The obtained metrics are averaged over $10$ different random splits of train, validation and test sets. For the MovieLens10M and the Netflix datasets, $20$\% of cold-start items are sampled and they are split into the cold validation and test sets. Then we proceed with a random $80\%$, $10\%$, $10\%$ split on the rest of the click matrix to get the train, warm validation and test sets. For the MovieLens10 dataset, the hybrid validation and test sets are obtained after combining the cold and warm validation and test sets. All models are trained on the training set with hyper-parameter optimization performed on the validation set. 

Regarding item metadata, we take the same set of attributes used in~\cite{bernardis2021nfc,CLCRec}. For Amazon Video Games, we use {\it title name, description, feature description, brand} and {\it categories}. For MovieLens Hetrec, we use {\it movie years, genres, actors, directors, countries,} and {\it locations}. For MovieLens10M, we use {\it title, description, genres, actors,} and {\it directors}. For Netflix, we use {\it title, genres, actors, directors, producers,} and {\it composers}. For the Netflix dataset, we join the titles of the Netflix prize dataset with the titles of the IMDb non-commercial dataset~\cite{imdb}.

\begin {table}[h!]
	\begin{center}
\small
			\begin{tabular}{c| c c c}
				&\# users &\# items & Interactions (\%)\\ [0.5ex] 
				\hline\hline
				Pinterest~\cite{he2017neural}&55187 & 9916&0.27\\ 
								\hline
				MovieLens1M~\cite{he2017neural}&6040&3706&4.47\\
				\hline
		\end{tabular}
	\end{center} \caption{Characteristics of the warm data splits used in experiments ordered by sparsity level.}\label{wdatasets}
	\end{table}

	\paragraph{Warm splits} We also present our analysis on publicly available data splits\footnote{\url{https://github.com/hexiangnan/neural_collaborative_filtering}} for warm-start recommendation that received a lot of attention in the literature~\cite{he2017neural,dacrema2019we,rendle2020neural}. The dataset characteristics are summarized in Table~\ref{wdatasets}. The splits were generated by selecting users with at least $20$ ratings and using the leave-one-out evaluation, ranking the last click together with $100$ randomly drawn negative samples. The validation set is generated from the training set using the same protocol as in~\cite{rendle2020neural}. 


	
	\subsection{Metrics and statistical significance} We leverage standard metrics reported in recommendation literature~\cite{liang2018variational} namely, hit rate (hr@k) and normalized 
	discounted cumulative gain (ndcg@k). Given a validation or test set containing a set of users $\mathcal{U}'\subseteq\mathcal{U}$, let $\mathcal{I}_u \subseteq \mathcal{I}$ be the set of items clicked by user $u$. 
	The metric hr@k is defined as:
	\begin{equation}
	\text{hr@k} = \frac{1}{|\mathcal{U}'|} \sum_{u\in\mathcal{U}'} \frac{\sum_{i\in\mathcal{I}_u}\mathbb{I}(r_{ui}\leq k)}{\min(k,\left|\mathcal{I}_u\right|)},
	\end{equation}
	
	where $\mathbb{I}$ is the indicator function 
	and $r_{ui}$ the predicted rank of item $i$ for user $u$. The ndcg@k is defined as:
	\begin{equation}
	\text{ndcg@k} = \frac{1}{|\mathcal{U}'|}\sum_{u\in\mathcal{U}'} \left(\sum_{i\in\mathcal{I}_u}\frac{\mathbb{I}(r_{ui}\leq k)}{\log_2 (1+r_{ui})}\bigg/\sum_{j\in K'}\frac{1}{\log_2(1+j)}\right),
	\label{metrics}
	\end{equation}
	where the set $K'$ is defined as $K'=\{1,\ldots,\min(k,|\mathcal{I}_u|)\}$. 
 
In the rest of the paper, we report average gain observed across all users in the test datasets. We also performed bootstrapping analysis to assess the significance of our results by sampling 20\% of the users in each test dataset and repeating the process 500 times, in order to compute 95\% confidence intervals, and we indicate when our results achieve statistical significance. 

\subsection{Results for Cold Splits} 
We now report ndcg@k and hr@k metrics on the cold data splits. We first focus on incorporating item features with the same featurization policy as in~\cite{bernardis2021nfc,CLCRec}. A study into more sophisticated ways to encode the item metadata features is also conducted. We discuss embeddings and alignment strategies to achieve the best results with item metadata features only. Finally, we explore the incorporation of more item metadata features and discuss the lifts obtained. Regarding the baselines considered, we largely make use of the code repository\footnote{\url{https://github. com/cesarebernardis/NeuralFeatureCombiner}} of~\cite{bernardis2021nfc} for hyper-parameter tuning, training and scoring. Hyper-parameter optimization was done via grid search on the validation sets, following the exact same process across all models. We train the baselines and our proposed approach on an ml.p3.2xlarge EC2 instance. 

\par{\bf Baselines.} We report the results for several competitive baselines: NFC~\cite{bernardis2021nfc}, ItemKNNCF~\cite{lops2011content}, Wide and Deep~\cite{cheng2016wide}, and FM~\cite{rendle2010factorization}. NFC is the hybrid model of~\cite{bernardis2021nfc}, which was reported to largely beat SOTA for cold item recommendations. ItemKNNCF is an item content-based k-nearest neighbour approach, which was proven to be competitive across a variety of tasks despite its simplicity~\cite{dacrema2019we}. Wide and Deep consists of training jointly wide linear models and deep neural networks to combine the benefits of memorization and generalization for recommender systems. Finally, Factorization Machines (FM) is a classic competitive baselines when dealing with cold-start recommendations~\cite{pan2019warm} and explicitly model first and second order interactions of feature and click embeddings. We also experimented with CLCRec~\cite{CLCRec}, and its extension EQUAL~\cite{equal}. Finally, we experimented with CVAR~\cite{zhao2022improving}, however the code repository of~\cite{zhao2022improving} did not scale to the size of our datasets.


\begin {table}[h!]
\begin{center}
\resizebox{\columnwidth}{!}{
	\begin{tabular}{c| c c c c }
		&\multicolumn{4}{c}{\textbf{MovieLens Hetrec}}\\
		
		& hr@10 &ndcg@10 & hr@25 &ndcg@25 \\ [0.5ex] 
		\hline
		NFC& \underline{0.1904} &\underline{0.2076} & 0.1748&\underline{0.1866} \\ 
		\hline
		ItemKNNCF&0.1175  &0.1335 &0.1130 & 0.1214\\
		\hline
		Wide and Deep&0.1555 &0.1762&0.1479& 0.1588 \\
		\hline
		FM& 0.1790 &0.1946 &\underline{0.1808} &0.1842\\
		\hline
            CLCRec & 0.0815  &0.0763 &0.0909 &0.0848\\
            \hline
		EQUAL & 0.1310 &0.1470 &0.1124 &0.1252\\
  \hline
		\manet~(Eq~\ref{genericEASE})&\textbf{0.2928}  &\textbf{0.3071} & \textbf{0.2717}&\textbf{0.2826}\\
		\hline
		Lift& +53.8\%$^{*}$ &+47.9\%$^{*}$&+49.7\%$^{*}$ &+51.4\%$^{*}$\\
		\hline\hline
  		&\multicolumn{4}{c}{\textbf{Amazon Video Games}} \\
		
		&hr@10 &ndcg@10 & hr@25 &ndcg@25\\ [0.5ex] 
		\hline
		NFC& \underline{0.1231} &0.0785  & \underline{0.2099} &\underline{0.1040}\\ 
		\hline
		ItemKNNCF& 0.1187 &\underline{0.0786} &0.1842&0.0980\\
		\hline
		Wide and Deep& 0.0103&0.0060& 0.0200&0.0088 \\
		\hline
		FM& 0.0080  &0.0134 &0.0129 &0.0302\\
            \hline
		CLCRec & 0.1031  &0.0676 &0.1804 &0.0903\\
            \hline
		EQUAL & 0.1171  &0.0778 &0.1957 &0.1007\\
            \hline
		\manet~(Eq~\ref{genericEASE})&\textbf{0.1568} & \textbf{0.1047}& \textbf{0.2433}&\textbf{0.1305}\\
		\hline
	    Lift& +27.4\%$^{*}$ &+33.2\%$^{*}$&+15.9\%$^{*}$ &+25.5\%$^{*}$\\
		\hline\hline
&\multicolumn{4}{c}{\textbf{MovieLens10M}}\\
		
		& hr@10& ndcg@10 & hr@25 &ndcg@25\\ [0.5ex] 
		\hline
		NFC&  0.0322 & 0.0379&0.0342 & 0.0366   \\ 
		\hline
		ItemKNNCF& \underline{0.0961}&  \underline{0.0865}& \underline{0.1525}& \underline{0.1054}     \\
		\hline
		Wide and Deep& 0.0576 & 0.0571&0.0864 &  0.0654    \\
		\hline
		FM& 0.0718 & 0.0634& 0.1064& 0.0737   \\
            \hline
            CLCRec& 0.0752& 0.0443& 0.1102 & 0.0851 \\
		\hline
              EQUAL& 0.0864& 0.0626 & 0.1309&0.1010 \\
		\hline
		\manet~(Eq~\ref{genericEASE})& \textbf{0.1094}& \textbf{0.0938}&\textbf{0.1956} &\textbf{0.1236}  \\
		\hline
		Lift& +13.8\%$^{*}$ &+8.4\%$^{*}$&+28.3\%$^{*}$ &+17.3\%$^{*}$\\
		\hline\hline 
  		&\multicolumn{4}{c}{\textbf{Netflix}}\\
		
		& hr@10& ndcg@10 & hr@25 &ndcg@25\\ [0.5ex] 
		\hline
		NFC&  \underline{0.1087} & 0.1098& \underline{0.1278}&  0.1114 \\ 
		\hline
		ItemKNNCF& 0.1081& \underline{0.1115} &0.1249&  \underline{0.1127}  \\
		\hline
		Wide and Deep& 0.0188& 0.0146&0.0250 &  0.0180    \\
		\hline
		FM& 0.0586& 0.0483&0.0990 &  0.0651 \\
            \hline
            CLCRec& 0.0934& 0.0987&0.1123 & 0.1074 \\
		\hline
          EQUAL& 0.1013& 0.1027&0.1252 & 0.1109 \\
		\hline
		\manet~(Eq~\ref{genericEASE})&\textbf{0.1369}& \textbf{0.1412}& \textbf{0.1574}&\textbf{0.1411}\\
		\hline
		Lift&+25.9\%$^{*}$  &+26.6\%$^{*}$&+23.2\%$^{*}$ &+25.2\%$^{*}$\\
		\hline
		
\end{tabular}}
\end{center}
\caption{Baseline comparison of \manet\ across benchmarking datasets on cold-start  recommendation. Top performing and second best scores are respectively in bold and underlined. Lift is reported between \manet\ and best performing baseline and $^{*}$ indicates statistical significance.}
\label{tab:hetrec}
\end{table}

\paragraph{\bf Comparison with baselines} For fair comparison, we adopt the same featurization strategies reported in~\cite{bernardis2021nfc,CLCRec} to produce the feature matrix $\textbf{F}$, that is, the text features are embedded through tf-idf with $1000$ as vocabulary size and the categorical features through multi-hot encoding. For the non-negative decreasing function $h$ used in the regularization term $\mathbf{D^{\text{R}}}$ of equation~\eqref{ali1}, we use a step linear decreasing function. We take the $10^{\text{th}}$ percentile $p$ of the distribution of the number of ratings per item $r^{(i)}=\sum_i X_{.i}$ and  the linear function is written $h\left(r^{(i)}\right)=k\left(p-r^{(i)}\right)\ \text{if}\ r^{(i)}\leq p,\ h\left(r^{(i)}\right)=0\ \text{if}\ r^{(i)}\geq p$, with $k = \beta/p$. We keep $f^E_{\phi}=I$ for the embedding model and for the alignment model $f^A_{\zeta}$ we used equation~\eqref{weighted_cosine}, and optimize for the first and second order scalar terms. For the backbone model $f_{\theta}^B$, we used equation~\eqref{genericEASE} which is the fastest to run.

Table~\ref{tab:hetrec} shows the lifts obtained by~\manet~on the MovieLens Hetrec, Amazon Video Games, MovieLens10M, and Netflix datasets, all with significant margins. Hyper-parameter optimization was performed on validation sets for all models via grid search. Regarding \manet, the hyper-parameters obtained on the MovieLens Hetrec validation sets are $\delta =20$ (equation~\ref{weighted_cosine}), $\lambda_1=1$ (equation~\ref{genericEASE}), $\alpha=1$ (equation~\ref{ali1}), $\beta=100$ (term $\mathbf{D^{\text{R}}}$ in equation~\ref{ali1}). For the Amazon Video Games dataset, the hyperparameters obtained on the validation sets are $\delta =50$, $\lambda_1=1$, $\alpha=10$, $\beta=100$. We note that learning the first and second order coefficients in the alignment function of~\eqref{weighted_cosine} did not lead to improvements, which we relate to the high sparsity observed in the data (click percentage of 0.06\%, see Table~\ref{cdatasets}). For the MovieLens10M dataset, the hyperparameters obtained on the validation set are $\delta =50$, $\lambda_1=700$, $\alpha=1$, $\beta=60$. Finally, for the Netflix dataset, the hyperparameters obtained on the validation set are $\delta =100$, $\lambda_1=500$, $\alpha=1$, $\beta=100$.

\subsection{Ablation Studies}
Here we focus on MovieLens10M dataset although similar observation was made on other datasets as well and those results are not presented for sake of brevity. 

\paragraph{\bf Discussion on the encoding and embeddings functions} We have showed that our proposed approach can outperform competitive baselines on a variety of cold-start dataset splits. This can be further improved by utilising better embeddings and leveraging more features. As a result, we are first interested in the recommendation quality we get without the fusion term, by only considering item metadata. 

\begin {table}[h!]
\begin{center}
\resizebox{\columnwidth}{!}{
	\begin{tabular}{c| c c c c }
		&\multicolumn{4}{c}{\textbf{MovieLens10M}}\\
		
		& hr@10&hr@25& hr@50 &hr@100\\ [0.5ex] 

		\hline
		ItemKNN &0.0071 &0.0137 &0.0241 & 0.0449  \\
		\hline
		\manet~(Eq~\ref{weighted_cosine}, 1st order)&0.0326&0.0484&0.0757& 0.1235 \\
		\hline
		\manet~(Eq~\ref{weighted_cosine})& 0.0329& 0.0487&0.0776& 0.1302 \\
		\hline
		\manet~(Eq~\ref{weighted_cosine}, Falcon7B)&\underline{0.0373} & \underline{0.0612}&\underline{0.1007}&\underline{0.1545}   \\
		\hline
		\manet~(Eq~\ref{weighted_cosine}, Falcon7B (all))& 0.0344&0.0541&0.0905 & 0.1448  \\
		\hline
		\manet~(Siamese, Falcon7B (all))&0.0357& 0.0537& 0.0820& 0.1280  \\
		\hline
				\manet~(Eq~\ref{weighted_cosine}, ST)&\textbf{0.0378} &\textbf{0.0626} &\textbf{0.1028}& \textbf{0.1602} \\
		\hline
\end{tabular}}
\end{center}
\caption{Ablation study on the encoding and embeddings functions for metadata-based recommendations on the MovieLens10M dataset split. Top performing and second best scores are respectively in bold and underlined.}
\label{tab:metadata}
\end{table}

Table~\ref{tab:metadata} presents the results of different encoding, and embeddings strategies to get to metadata-based recommendations. The first step we tried was to replace the text tf-idf embeddings with pretrained Large Language Models (LLMs) embeddings: we experimented with Falcon-7B~\cite{almazrouei2023falcon} and Sentence Transformers (ST)~\cite{reimers-2019-sentence-bert} embeddings. We also experimented with learning the first order coefficients $\mu_i,i\in\{1,\cdots,N\}$ vs learning first and second order coefficients $\mu_i,\mu_{ij}$, with $i, j\in\{1,\cdots,N\}, j>i$ of equation~\eqref{weighted_cosine}. First, we notice that LLM pre-trained embeddings largely outperformed the tf-idf embeddings for hr@k and ndcg@k, for instance ST embeddings led to an increase of +22.8\% for hr@100 with respect to tf-idf embeddings. Second, we notice that learning scalar weights for second order terms, as per equation~\eqref{weighted_cosine}, brings a clear gain in ranking metrics. The second and third rows of Table~\ref{tab:metadata} show a consistent benefit over ranking metrics and a gain of 5.4\% for hr@100. This is because the multiplication of cosine similarities enable to model nonlinear interactions between sub-feature spaces. 

Finally, we also report results on converting the categorical features (actors, directors, genres) into numeric representations with Falcon7B, indicated as "Falcon7B (all)" in Table~\ref{tab:metadata}. This led to a small decrease in performance over using multi-hot encoded features, e.g., -6.8\% for hr@100. However this helped when training the Siamese network, and the best results on cold-start splits were reported using Falcon embeddings across all features. For the Siamese network, we used 1 projection layer followed by 1 cross attention layer between text and categorical features and 2 dense layers. We noticed that feeding categorical features to the Siamese network was leading the network to learn id associations between multi-hot encoded features and items, instead of semantic ones. The results we report here show that the Siamese network fall short of the best reported results, e.g., -6.6\% for recall@10.

\paragraph{\bf Fusion performance, and comparison of backbone models} The previously presented results give an indication of the performance of the fusion approaches (equations~\ref{genericVAE} and~\ref{genericEASE}), that we present in Table~\ref{tab:ensemble}. 

\begin {table}[h!]
\begin{center}
\resizebox{\columnwidth}{!}{
	\begin{tabular}{c| c c c c }
		&\multicolumn{4}{c}{\textbf{MovieLens10M}}\\
		& hr@10&hr@25& hr@50 &hr@100\\ [0.5ex] 
  		\hline 
  	\manet~(Eq~\ref{genericEASE}, ST)& \textbf{0.1385}&\textbf{0.2240}&0.3286 & 0.4692  \\
   \hline
		\manet~(Eq~\ref{genericEASE}, BoW)&0.1094 &0.1956 &0.2866&0.4203  \\
		\hline
			\manet~(Eq~\ref{genericEASE}, Siamese)& 0.0910&0.1601 &0.2457 &0.4109  \\
			\hline
  \hline
		\manet~(Eq~\ref{genericVAE}, ST)&0.0779& 0.2037 &\textbf{0.3831}&  \textbf{0.5072} \\
  \hline
  	\manet~(Eq~\ref{genericEASE}, ST)& \textbf{0.1385}&\textbf{0.2240}&0.3286 & 0.4692 \\
  \hline
  		\manet~(Eq~\ref{wls}, ST)&\underline{0.1282}& \underline{0.2204} &\underline{0.3421}& \underline{0.4789} \\
		\hline

\end{tabular}}
\end{center}
\caption{Ablation study on the MovieLens10M dataset for difference fusion strategies, and for different backbone models. Top performing and second best scores are respectively in bold and underlined. }
\label{tab:ensemble}
\end{table}

First, the Siamese approach which was trailing best results on the metadata-based learning task is also trailing after fusion. It falls behind the alignment function leveraging first and second order cosine similarities, as per equation~\eqref{weighted_cosine}, by 2.3\% on recall@100. As future work, we will explore improving the joint representation learning of item metadata with the Siamese network, by e.g., including external data to further help with the adaptation of the pre-trained embeddings. Second, we report the performance of our network with the 3 backbone models of equations~\eqref{genericVAE},~\eqref{genericEASE}, and~\eqref{wls}. The multi-VAE network was trained with one encoding and one decoding layer similarly to~\cite{liang2018variational} and after hyper-parameter turning, we trained the network over $20$ epochs, with a embedding dimension of $200$, and $D^{\text{A}}$ in equation~\eqref{ali1} chosen such that $\beta=50$. Our observation is that equation~\eqref{genericVAE} struggles with ranking metrics for low values of $k$, falling -$44.8\%$ and -$9.1\%$ behind~\eqref{genericEASE} for hr@10 and hr@25, while outperforming~\eqref{genericEASE} and~\eqref{wls} for high values of $k$, with a gain of +$16.6\%$ for hr@50 and +8.1\% for hr@100. We interpret that this is caused by the needs for deeper representations with more than one hidden layer, combined with the inaccuracy of the sampling Gaussian prior. We note that this difference in performance between the two backbone models was also appreciated in~\cite{vanvcura2021deep}, which focused on warm conditions, and where a late fusion was proposed to ensemble a multi-VAE branch and a neural EASE branch, by simply multiplying the softmax probabilities obtained. As future work, we may experiment with similar ensembling techniques to bring the best out of the backbone models, see e.g.~\cite{kim2019enhancing}.

\paragraph{\bf Adding images, and tags} We explore the incorporation of additional metadata into our framework. Table~\ref{tab:ablation} presents the results when we incorporated image embeddings and item tags to the metadata signals. 

\begin {table}[h!]
\begin{center}
\resizebox{\columnwidth}{!}{
	\begin{tabular}{c| c c c c }
		&\multicolumn{4}{c}{\textbf{MovieLens10M}}\\
		& hr@10&hr@25& hr@50 &hr@100\\ [0.5ex] 
		\hline
			\manet~(Eq~\ref{genericEASE}, ST) &0.1385&0.2240&0.3286 & 0.4692  \\
			\hline
	\manet~(Eq~\ref{genericEASE}, ST + images)& 0.1401&0.2267& 0.3315& 0.4721 \\
		\hline
				\manet~(Eq~\ref{genericEASE}, ST + images, tags)& 0.2194&0.3540 & 0.4808& 0.6168 \\
			\hline
\end{tabular}}
\end{center}
\caption{Results for the cold-start MovieLens10M dataset split on introducing additional metadata to our framework. }
\label{tab:ablation}
\end{table}

The MovieLens10M images were pulled from the Kaggle poster repository\footnote{https://www.kaggle.com/datasets/ghrzarea/movielens-20m-posters-for-machine-learning?select=MLP-20M}. We used the a pretrained CLIP model (ViT-H/14 - LAION-2B\footnote{laion/CLIP-ViT-H-14-laion2B-s32B-b79K}) to get image embeddings. We relied on this model as we also tried to generate keywords and captions from these images by leveraging the BLIP2-2.7b architecture but this did not lead to improvements in ranking metrics. On the contrary, Table~\ref{tab:ablation} shows clear benefits in adding the image embeddings to the framework, with gains ranging from 0.5\% to 1.4\% on the recall@k metrics. Finally, we also made use of the MovieLens tags provided as part of the MovieLens dataset and we simply encoded them through multi-hot encoding, after merging similar tags together, which led to a vocabulary of 2504 words. This is an informative feature, e.g., the movie "Dumb and Dumber" would come with the following tags: "Jim Carrey", "Jeff Daniel", "stupid", "comedy", "infantil", "hilarious". 

The inclusion of such tags led to a very significant additional improvement in ranking metrics, e.g. +30.7\% for hr@100. We also get to an impressive cold-start performance of 0.6168 for hr@100, which after checking is only 15.1\% behind the hr@100 we get on the warm split. In general, the inclusion of semantic features, i.e. LLM embeddings, image embeddings, and tags metadata, have led to a gain of 105.5\% on hr@10 and +46.7\% on hr@100 with respect to training the same model with BoW representation. Under the MovieLens set-up, tags are coming from user entries, however, we believe the process of tagging items could be repeated by a ML system utilising the world knowledge hence removing the needs for an user input. We leave this as future work.

\paragraph{\bf Results for transitioning from cold to warm conditions} Finally, we report the capability of the proposed approach to provide relevant recommendations for situations in which there is a mix of cold and warm items.  
	\begin {table}[h!]
	\begin{center}
		\resizebox{\columnwidth}{!}{
			\begin{tabular}{c| c c c c c c }
				&\multicolumn{6}{c}{\textbf{MovieLens10M}} \\
				&\multicolumn{2}{c}{\textbf{Warm}}&\multicolumn{2}{c}{\textbf{Cold}}&\multicolumn{2}{c}{\textbf{All}}\\
				& hr@10 &ncdg@10 & hr@10 &ndcg@10&hr@10 &ndcg@10\\ [0.5ex] 
				\hline\hline
				NFC&0.0945 & 0.0736 &0.0322&0.0379&0.0833&0.0767\\ 
				\hline
				ItemKNNCF& 0.0499 &0.0476&\underline{0.0962} &\underline{0.0865}&0.0509&0.0565\\
				\hline
    				Wide and Deep&0.1204&0.0907& 0.0576&0.0571&0.0879&0.0752\\
				\hline
        				FM&0.1461&0.1206&0.0718&0.0634&0.1012&0.0947\\
                    \hline
        				CLCRec&0.3157&0.2364&0.0752&0.0443&0.2335 &0.1979\\
                    \hline
        				EQUAL&\underline{0.3368}&\underline{0.2471}&0.0864&0.0626&\underline{0.2427}&\underline{0.2056}\\
				\hline
				\manet~(Eq~\ref{weighted_cosine}, ST)&0.0558&0.0494&0.0378&0.0384&0.0618 &0.0662\\
				\hline
				\manet~(Eq~\ref{genericEASE}, ST)&\textbf{0.3513} &\textbf{0.2890}&\textbf{0.1385}&\textbf{0.1150}&\textbf{0.2443} &\textbf{0.2283}\\
				\hline
		\end{tabular}}
	\end{center}
	\caption{Results on the MovieLens10M dataset under different evaluation scenarios with warm, mixed, and cold items. Top performing and second best scores are respectively in bold and underlined.}
	\label{tab:mix}
\end{table}

 Table~\ref{tab:mix} presents the results in the 3 test scenarios: warm items only, cold and warm items, and cold items only, against the 2 best performing approaches reported in~\cite{CLCRec}. We have seen that \manet~was largely beating SOTA results on cold-start splits and here we confirm its efficacy for warm and mixed situations. As for the hyperparameters, we get $\lambda_1=700$, $\alpha=1$, $\beta=0.5$, for the mixed split and $\lambda_1=500$, $\alpha=1$, $\beta=0$, for the warm split. For the mixed and warm situations, it also beats the baselines, with CLCRec and EQUAL following more closely. We repeat that our approach transitions to the backbone model in warm set ups, which can be selected from top performing collaborative filtering algorithms. 

\paragraph{\bf A note on time complexity} We note that the training process can be done either in sequential steps by first learning the best metadata representations (equations~\ref{cosine},~\ref{weighted_cosine}) and then training the backbone and fusion part of the network (equations~\ref{genericVAE},~\ref{genericEASE},~\ref{genericMSLIM}), or in an end-to-end fashion. The sequential implementation enables not to pass all the metadata feature matrix $\mathbf{F}$ at test time but rather the user specific generated embeddings. We report the training times of the approaches: on MovieLens10M, 0.43~s for ItemKNNCF, \num{5.3e3} sec for Wide and Deep, \num{1.3e3}~s for FM, \num{1.3e4}~s for NFC, \num{6.3e4}~s for EQUAL, 230~s for~\eqref{genericVAE}, and 34~s for~\eqref{genericEASE}; and on Amazon Video Games, 2~s for ItemKNNCF, 190~s for Wide and Deep, 27~s for FM, \num{8.2e3}~s for NFC, \num{1.8e3}~s for EQUAL, 440~s for equation~\eqref{genericVAE}, and 92~s for equation~\eqref{genericEASE}. The gain in training time over the second best performing approach is of one order of magnitude for the VAE backbone and two orders of magnitude for the EASE backbone. We note that the matrix implementation of~\eqref{genericEASE} will suffer the curse of dimensionality in the user space, however equation~\eqref{genericVAE} scales well as it trains across batches of users.

\subsection{Results for Warm Splits} 

We also put a focus on warm dataset splits to evaluate the merit of our proposed optimization strategy of equation~\eqref{wls}, and its closed form solution presented in equation~\eqref{genericMSLIM}. As for the weight matrices we use $w_0 = 1$ for
positive labels and tunable parameter $w_1$ for negative labels. Hence
the 3 parameters to optimise are $w_1$, $\lambda_1$, and $\gamma_1$. For the weight matrices we use $w_0=1$ for positive labels and tunable parameter $w_1$ for negative labels. Hence the 3 parameters to optimise are $w_1, \lambda_1$, and $\gamma_1$. Hyperparameter
optimization is done via grid search on the validation set optimising
for both ndcg@10 and recall@10, where the grids are defined with
parameters such as $w_1 \in \{0, 0.5, . . . , 1\}$, $\lambda_1 \in \{0, 100, . . . , 2000\}$,
and $\gamma_1 \in \{0, 100, . . . , 2000\}$. We get $w_1$ = 0.25, $\lambda_1 = 300$, $\gamma_1 = 800$
for MovieLens1M, $w_1 = 0.05$, $\lambda_1 = 300$, $\gamma_1 = 100$ for Pinterest. We implement our approach with a batch array job with on-demand AWS Fargate resources and 1 vCPU of 8 GB of memory per job, which runs in 1 min for the MovieLens dataset and 7 min for the Pinterest dataset. 

\begin {table}[h!]
\begin{center}
\resizebox{\columnwidth}{!}{
\begin{tabular}{c| c c c c c}		&\multicolumn{2}{c}{\textbf{MovieLens1M}}&\multicolumn{2}{c}{\textbf{Pinterest}}& \\
		
		& hr@10 &ncdg@10 & hr@10 &ndcg@10\\ [0.5ex] 
		\hline\hline
		Popularity& 0.4535 & 0.2543 & 0.2740&0.1409\\ 
		\hline
		iALS& 0.711 & 0.4383 & 0.8762 &0.5590 \\
  \hline
		NeuMF& 0.7093&0.4039&\underline{0.8777}&0.5576\\
		\hline
		SLIM& 0.7162 & 0.4468& 0.8679 &0.5601\\
		\hline
		MF& \underline{0.7294} & \underline{0.4523} & \textbf{0.8895}&\textbf{0.5794}\\
		\hline
		EASE& 0.7184 & 0.4494 & 0.8634 & 0.5616 \\
		\hline
		\manet~(Eq~\ref{wls})& \textbf{0.7301} & \textbf{0.4560} & 0.8748 &\underline{0.5683}  \\
		\hline
		Lift &+0.1\%  & +0.8\%$^{*}$&-1.5\%  &-1.9\%$^{*}$   \\
		\hline
\end{tabular}}
\end{center}
\caption{Baseline comparison of \manet\ on warm-start benchmarking datasets. Top performing and second best scores are respectively in bold and underlined. Lift is reported between \manet\ and best performing baseline and $^{*}$ indicates statistical significance.}
\label{tab:table1}
\end{table}

 In Table~\ref{tab:table1}, we report the competitive baselines already reported in~\cite{rendle2020neural}, to which we add the EASE model~\cite{steck2019embarrassingly} and our optimization strategy~\eqref{genericMSLIM}. The baselines are the neural matrix factorization (NMF) approach of~\cite{he2017neural}, the implicit alternating least squares (iALS) collaborative filtering method~\cite{rendle2022revisiting}, the original SLIM model~\cite{ning2011slim}, and the matrix factorization (MF) implementation of~\cite{rendle2020neural} which is the reported SOTA result on the 2 datasets. It is interesting to note that weighted least squares beats matrix
factorization for the least sparse dataset (average margin +0.5\%) and
fails to beat it for the sparse dataset (average margin -1.7\%). Another remark is
that we consistently beat SLIM (+2.0\%, +1.1\%), and EASE
(+1.5\%, +1.3\%), meaning that our closed-form weighted least squares should always
be considered as strong baseline.

\section{Conclusion}
We introduced \manet, a novel algorithm to enhance cold start recommendations by integrating semantic information. This was achieved by combining similarities derived from user click actions and metadata features using a regularization term, aligning cold-start items with the collaborative filtering space. 
Experimental results demonstrated that \manet\ significantly outperformed reported SOTA results on cold-start datasets with varying sparsity and metadata characteristics, showing improvements ranging from +8.4\% to +53.8\% on hr@k and ndcg@k metrics while being competitive on warm-start recommendation datasets.
Moreover, \manet\ achieved these results while being one order of magnitude faster to train compared to the second-best performing baseline. The inclusion and utilization of semantic features led to substantial gains ranging from +46.8\% to +105.5\% compared to Bag-of-Words (BoW) representations. 
Future work will focus on refining joint metadata embeddings by incorporating external data sources, exploring generative models for informative keyword generation, and investigating more sophisticated end-to-end training and merging mechanisms within the framework.

\bibliographystyle{ACM-Reference-Format}
\bibliography{main-acm-recsys}


\begin{thebibliography}{50}


\ifx \showCODEN    \undefined \def \showCODEN     #1{\unskip}     \fi
\ifx \showDOI      \undefined \def \showDOI       #1{#1}\fi
\ifx \showISBNx    \undefined \def \showISBNx     #1{\unskip}     \fi
\ifx \showISBNxiii \undefined \def \showISBNxiii  #1{\unskip}     \fi
\ifx \showISSN     \undefined \def \showISSN      #1{\unskip}     \fi
\ifx \showLCCN     \undefined \def \showLCCN      #1{\unskip}     \fi
\ifx \shownote     \undefined \def \shownote      #1{#1}          \fi
\ifx \showarticletitle \undefined \def \showarticletitle #1{#1}   \fi
\ifx \showURL      \undefined \def \showURL       {\relax}        \fi
\providecommand\bibfield[2]{#2}
\providecommand\bibinfo[2]{#2}
\providecommand\natexlab[1]{#1}
\providecommand\showeprint[2][]{arXiv:#2}

\bibitem[imd({[n.\,d.]})]%
        {imdb}
 \bibinfo{year}{[n.\,d.]}\natexlab{}.
\newblock \bibinfo{title}{IMDb Non-Commercial Datasets}.
\newblock
  \bibinfo{howpublished}{\url{https://developer.imdb.com/non-commercial-datasets/}}.
\newblock


\bibitem[ml1({[n.\,d.]})]%
        {ml10m}
 \bibinfo{year}{[n.\,d.]}\natexlab{}.
\newblock \bibinfo{title}{MovieLens 10M Dataset}.
\newblock
  \bibinfo{howpublished}{\url{https://grouplens.org/datasets/movielens/10m/}}.
\newblock


\bibitem[net({[n.\,d.]})]%
        {netflix}
 \bibinfo{year}{[n.\,d.]}\natexlab{}.
\newblock \bibinfo{title}{Netflix Prize data}.
\newblock
  \bibinfo{howpublished}{\url{https://www.kaggle.com/datasets/netflix-inc/netflix-prize-data}}.
\newblock


\bibitem[Almazrouei et~al\mbox{.}(2023)]%
        {almazrouei2023falcon}
\bibfield{author}{\bibinfo{person}{Ebtesam Almazrouei}, \bibinfo{person}{Hamza
  Alobeidli}, \bibinfo{person}{Abdulaziz Alshamsi}, \bibinfo{person}{Alessandro
  Cappelli}, \bibinfo{person}{Ruxandra Cojocaru}, \bibinfo{person}{Mérouane
  Debbah}, \bibinfo{person}{Étienne Goffinet}, \bibinfo{person}{Daniel
  Hesslow}, \bibinfo{person}{Julien Launay}, \bibinfo{person}{Quentin
  Malartic}, \bibinfo{person}{Daniele Mazzotta}, \bibinfo{person}{Badreddine
  Noune}, \bibinfo{person}{Baptiste Pannier}, {and} \bibinfo{person}{Guilherme
  Penedo}.} \bibinfo{year}{2023}\natexlab{}.
\newblock \bibinfo{title}{The Falcon Series of Open Language Models}.
\newblock
\newblock
\showeprint[arxiv]{2311.16867}~[cs.CL]


\bibitem[Bai and Fan(2017)]%
        {bai2017incorporating}
\bibfield{author}{\bibinfo{person}{Bing Bai} {and} \bibinfo{person}{Yushun
  Fan}.} \bibinfo{year}{2017}\natexlab{}.
\newblock \showarticletitle{Incorporating field-aware deep embedding networks
  and gradient boosting decision trees for music recommendation}. In
  \bibinfo{booktitle}{\emph{The 11th ACM International Conference on Web Search
  and Data Mining (WSDM). ACM, London, England}}, Vol.~\bibinfo{volume}{7}.
\newblock


\bibitem[Bernardis and Cremonesi(2021)]%
        {bernardis2021nfc}
\bibfield{author}{\bibinfo{person}{Cesare Bernardis} {and}
  \bibinfo{person}{Paolo Cremonesi}.} \bibinfo{year}{2021}\natexlab{}.
\newblock \showarticletitle{NFC: a deep and hybrid item-based model for item
  cold-start recommendation}.
\newblock \bibinfo{journal}{\emph{User Modeling and User-Adapted Interaction}}
  (\bibinfo{year}{2021}), \bibinfo{pages}{1--34}.
\newblock


\bibitem[Borgeaud et~al\mbox{.}(2022)]%
        {borgeaud2022improving}
\bibfield{author}{\bibinfo{person}{Sebastian Borgeaud}, \bibinfo{person}{Arthur
  Mensch}, \bibinfo{person}{Jordan Hoffmann}, \bibinfo{person}{Trevor Cai},
  \bibinfo{person}{Eliza Rutherford}, \bibinfo{person}{Katie Millican},
  \bibinfo{person}{George~Bm Van Den~Driessche}, \bibinfo{person}{Jean-Baptiste
  Lespiau}, \bibinfo{person}{Bogdan Damoc}, \bibinfo{person}{Aidan Clark},
  {et~al\mbox{.}}} \bibinfo{year}{2022}\natexlab{}.
\newblock \showarticletitle{Improving language models by retrieving from
  trillions of tokens}. In \bibinfo{booktitle}{\emph{International conference
  on machine learning}}. PMLR, \bibinfo{pages}{2206--2240}.
\newblock


\bibitem[Bromley et~al\mbox{.}(1993)]%
        {10.5555/2987189.2987282}
\bibfield{author}{\bibinfo{person}{Jane Bromley}, \bibinfo{person}{Isabelle
  Guyon}, \bibinfo{person}{Yann LeCun}, \bibinfo{person}{Eduard S\"{a}ckinger},
  {and} \bibinfo{person}{Roopak Shah}.} \bibinfo{year}{1993}\natexlab{}.
\newblock \showarticletitle{Signature Verification Using a "Siamese" Time Delay
  Neural Network}. In \bibinfo{booktitle}{\emph{Proceedings of the 6th
  International Conference on Neural Information Processing Systems}} (Denver,
  Colorado) \emph{(\bibinfo{series}{NIPS'93})}. \bibinfo{publisher}{Morgan
  Kaufmann Publishers Inc.}, \bibinfo{address}{San Francisco, CA, USA},
  \bibinfo{pages}{737–744}.
\newblock


\bibitem[Chen and Chen(2019)]%
        {chen2019differentiating}
\bibfield{author}{\bibinfo{person}{Hung-Hsuan Chen} {and} \bibinfo{person}{Pu
  Chen}.} \bibinfo{year}{2019}\natexlab{}.
\newblock \showarticletitle{Differentiating regularization weights--A simple
  mechanism to alleviate cold start in recommender systems}.
\newblock \bibinfo{journal}{\emph{ACM Transactions on Knowledge Discovery from
  Data (TKDD)}} \bibinfo{volume}{13}, \bibinfo{number}{1}
  (\bibinfo{year}{2019}), \bibinfo{pages}{1--22}.
\newblock


\bibitem[Cheng et~al\mbox{.}(2016)]%
        {cheng2016wide}
\bibfield{author}{\bibinfo{person}{Heng-Tze Cheng}, \bibinfo{person}{Levent
  Koc}, \bibinfo{person}{Jeremiah Harmsen}, \bibinfo{person}{Tal Shaked},
  \bibinfo{person}{Tushar Chandra}, \bibinfo{person}{Hrishi Aradhye},
  \bibinfo{person}{Glen Anderson}, \bibinfo{person}{Greg Corrado},
  \bibinfo{person}{Wei Chai}, \bibinfo{person}{Mustafa Ispir}, {et~al\mbox{.}}}
  \bibinfo{year}{2016}\natexlab{}.
\newblock \showarticletitle{Wide \& deep learning for recommender systems}. In
  \bibinfo{booktitle}{\emph{Proceedings of the 1st workshop on deep learning
  for recommender systems}}. \bibinfo{pages}{7--10}.
\newblock


\bibitem[Covington et~al\mbox{.}(2016)]%
        {covington2016deep}
\bibfield{author}{\bibinfo{person}{Paul Covington}, \bibinfo{person}{Jay
  Adams}, {and} \bibinfo{person}{Emre Sargin}.}
  \bibinfo{year}{2016}\natexlab{}.
\newblock \showarticletitle{Deep neural networks for youtube recommendations}.
  In \bibinfo{booktitle}{\emph{Proceedings of the 10th ACM conference on
  recommender systems}}. \bibinfo{pages}{191--198}.
\newblock


\bibitem[Dacrema et~al\mbox{.}(2019)]%
        {dacrema2019we}
\bibfield{author}{\bibinfo{person}{Maurizio~Ferrari Dacrema},
  \bibinfo{person}{Paolo Cremonesi}, {and} \bibinfo{person}{Dietmar Jannach}.}
  \bibinfo{year}{2019}\natexlab{}.
\newblock \showarticletitle{Are we really making much progress? A worrying
  analysis of recent neural recommendation approaches}. In
  \bibinfo{booktitle}{\emph{Proceedings of the 13th ACM Conference on
  Recommender Systems}}. \bibinfo{pages}{101--109}.
\newblock


\bibitem[Deshpande and Karypis(2004)]%
        {10.1145/963770.963776}
\bibfield{author}{\bibinfo{person}{Mukund Deshpande} {and}
  \bibinfo{person}{George Karypis}.} \bibinfo{year}{2004}\natexlab{}.
\newblock \showarticletitle{Item-Based Top-N Recommendation Algorithms}.
\newblock \bibinfo{journal}{\emph{ACM Trans. Inf. Syst.}} \bibinfo{volume}{22},
  \bibinfo{number}{1} (\bibinfo{date}{jan} \bibinfo{year}{2004}),
  \bibinfo{pages}{143–177}.
\newblock
\showISSN{1046-8188}
\urldef\tempurl%
\url{https://doi.org/10.1145/963770.963776}
\showDOI{\tempurl}


\bibitem[Du et~al\mbox{.}(2020)]%
        {du2020learn}
\bibfield{author}{\bibinfo{person}{Xiaoyu Du}, \bibinfo{person}{Xiang Wang},
  \bibinfo{person}{Xiangnan He}, \bibinfo{person}{Zechao Li},
  \bibinfo{person}{Jinhui Tang}, {and} \bibinfo{person}{Tat-Seng Chua}.}
  \bibinfo{year}{2020}\natexlab{}.
\newblock \showarticletitle{How to learn item representation for cold-start
  multimedia recommendation?}. In \bibinfo{booktitle}{\emph{Proceedings of the
  28th ACM International Conference on Multimedia}}.
  \bibinfo{pages}{3469--3477}.
\newblock


\bibitem[Gantner et~al\mbox{.}(2010)]%
        {gantner2010learning}
\bibfield{author}{\bibinfo{person}{Zeno Gantner}, \bibinfo{person}{Lucas
  Drumond}, \bibinfo{person}{Christoph Freudenthaler}, \bibinfo{person}{Steffen
  Rendle}, {and} \bibinfo{person}{Lars Schmidt-Thieme}.}
  \bibinfo{year}{2010}\natexlab{}.
\newblock \showarticletitle{Learning attribute-to-feature mappings for
  cold-start recommendations}. In \bibinfo{booktitle}{\emph{2010 IEEE
  international conference on data mining}}. IEEE, \bibinfo{pages}{176--185}.
\newblock


\bibitem[Geng et~al\mbox{.}(2015)]%
        {7410843}
\bibfield{author}{\bibinfo{person}{Xue Geng}, \bibinfo{person}{Hanwang Zhang},
  \bibinfo{person}{Jingwen Bian}, {and} \bibinfo{person}{Tat-Seng Chua}.}
  \bibinfo{year}{2015}\natexlab{}.
\newblock \showarticletitle{Learning Image and User Features for Recommendation
  in Social Networks}. In \bibinfo{booktitle}{\emph{2015 IEEE International
  Conference on Computer Vision (ICCV)}}. \bibinfo{pages}{4274--4282}.
\newblock
\urldef\tempurl%
\url{https://doi.org/10.1109/ICCV.2015.486}
\showDOI{\tempurl}


\bibitem[He et~al\mbox{.}(2017)]%
        {he2017neural}
\bibfield{author}{\bibinfo{person}{Xiangnan He}, \bibinfo{person}{Lizi Liao},
  \bibinfo{person}{Hanwang Zhang}, \bibinfo{person}{Liqiang Nie},
  \bibinfo{person}{Xia Hu}, {and} \bibinfo{person}{Tat-Seng Chua}.}
  \bibinfo{year}{2017}\natexlab{}.
\newblock \showarticletitle{Neural collaborative filtering}. In
  \bibinfo{booktitle}{\emph{Proceedings of the 26th international conference on
  world wide web}}. \bibinfo{pages}{173--182}.
\newblock


\bibitem[Hu et~al\mbox{.}(2008)]%
        {4781121}
\bibfield{author}{\bibinfo{person}{Yifan Hu}, \bibinfo{person}{Yehuda Koren},
  {and} \bibinfo{person}{Chris Volinsky}.} \bibinfo{year}{2008}\natexlab{}.
\newblock \showarticletitle{Collaborative Filtering for Implicit Feedback
  Datasets}. In \bibinfo{booktitle}{\emph{2008 Eighth IEEE International
  Conference on Data Mining}}. \bibinfo{pages}{263--272}.
\newblock
\urldef\tempurl%
\url{https://doi.org/10.1109/ICDM.2008.22}
\showDOI{\tempurl}


\bibitem[Khandelwal et~al\mbox{.}(2019)]%
        {khandelwal2019generalization}
\bibfield{author}{\bibinfo{person}{Urvashi Khandelwal}, \bibinfo{person}{Omer
  Levy}, \bibinfo{person}{Dan Jurafsky}, \bibinfo{person}{Luke Zettlemoyer},
  {and} \bibinfo{person}{Mike Lewis}.} \bibinfo{year}{2019}\natexlab{}.
\newblock \showarticletitle{Generalization through memorization: Nearest
  neighbor language models}.
\newblock \bibinfo{journal}{\emph{arXiv preprint arXiv:1911.00172}}
  (\bibinfo{year}{2019}).
\newblock


\bibitem[Kim and Suh(2019)]%
        {kim2019enhancing}
\bibfield{author}{\bibinfo{person}{Daeryong Kim} {and} \bibinfo{person}{Bongwon
  Suh}.} \bibinfo{year}{2019}\natexlab{}.
\newblock \showarticletitle{Enhancing VAEs for collaborative filtering:
  flexible priors \& gating mechanisms}. In
  \bibinfo{booktitle}{\emph{Proceedings of the 13th ACM Conference on
  Recommender Systems}}. \bibinfo{pages}{403--407}.
\newblock


\bibitem[Koren(2008)]%
        {koren2008factorization}
\bibfield{author}{\bibinfo{person}{Yehuda Koren}.}
  \bibinfo{year}{2008}\natexlab{}.
\newblock \showarticletitle{Factorization meets the neighborhood: a
  multifaceted collaborative filtering model}. In
  \bibinfo{booktitle}{\emph{Proceedings of the 14th ACM SIGKDD international
  conference on Knowledge discovery and data mining}}.
  \bibinfo{pages}{426--434}.
\newblock


\bibitem[Koren et~al\mbox{.}(2009)]%
        {koren2009matrix}
\bibfield{author}{\bibinfo{person}{Yehuda Koren}, \bibinfo{person}{Robert
  Bell}, {and} \bibinfo{person}{Chris Volinsky}.}
  \bibinfo{year}{2009}\natexlab{}.
\newblock \showarticletitle{Matrix factorization techniques for recommender
  systems}.
\newblock \bibinfo{journal}{\emph{Computer}} \bibinfo{volume}{42},
  \bibinfo{number}{8} (\bibinfo{year}{2009}), \bibinfo{pages}{30--37}.
\newblock


\bibitem[Lam et~al\mbox{.}(2008)]%
        {lam2008addressing}
\bibfield{author}{\bibinfo{person}{Xuan~Nhat Lam}, \bibinfo{person}{Thuc Vu},
  \bibinfo{person}{Trong~Duc Le}, {and} \bibinfo{person}{Anh~Duc Duong}.}
  \bibinfo{year}{2008}\natexlab{}.
\newblock \showarticletitle{Addressing cold-start problem in recommendation
  systems}. In \bibinfo{booktitle}{\emph{Proceedings of the 2nd international
  conference on Ubiquitous information management and communication}}.
  \bibinfo{pages}{208--211}.
\newblock


\bibitem[Liang et~al\mbox{.}(2018)]%
        {liang2018variational}
\bibfield{author}{\bibinfo{person}{Dawen Liang}, \bibinfo{person}{Rahul~G
  Krishnan}, \bibinfo{person}{Matthew~D Hoffman}, {and} \bibinfo{person}{Tony
  Jebara}.} \bibinfo{year}{2018}\natexlab{}.
\newblock \showarticletitle{Variational autoencoders for collaborative
  filtering}. In \bibinfo{booktitle}{\emph{Proceedings of the 2018 world wide
  web conference}}. \bibinfo{pages}{689--698}.
\newblock


\bibitem[Long et~al\mbox{.}(2022)]%
        {long2022retrieval}
\bibfield{author}{\bibinfo{person}{Alexander Long}, \bibinfo{person}{Wei Yin},
  \bibinfo{person}{Thalaiyasingam Ajanthan}, \bibinfo{person}{Vu Nguyen},
  \bibinfo{person}{Pulak Purkait}, \bibinfo{person}{Ravi Garg},
  \bibinfo{person}{Alan Blair}, \bibinfo{person}{Chunhua Shen}, {and}
  \bibinfo{person}{Anton van~den Hengel}.} \bibinfo{year}{2022}\natexlab{}.
\newblock \showarticletitle{Retrieval augmented classification for long-tail
  visual recognition}. In \bibinfo{booktitle}{\emph{Proceedings of the IEEE/CVF
  Conference on Computer Vision and Pattern Recognition}}.
  \bibinfo{pages}{6959--6969}.
\newblock


\bibitem[Lops et~al\mbox{.}(2011)]%
        {lops2011content}
\bibfield{author}{\bibinfo{person}{Pasquale Lops}, \bibinfo{person}{Marco
  De~Gemmis}, {and} \bibinfo{person}{Giovanni Semeraro}.}
  \bibinfo{year}{2011}\natexlab{}.
\newblock \showarticletitle{Content-based recommender systems: State of the art
  and trends}.
\newblock \bibinfo{journal}{\emph{Recommender systems handbook}}
  (\bibinfo{year}{2011}), \bibinfo{pages}{73--105}.
\newblock


\bibitem[McAuley et~al\mbox{.}(2015)]%
        {mcauley2015image}
\bibfield{author}{\bibinfo{person}{Julian McAuley},
  \bibinfo{person}{Christopher Targett}, \bibinfo{person}{Qinfeng Shi}, {and}
  \bibinfo{person}{Anton Van Den~Hengel}.} \bibinfo{year}{2015}\natexlab{}.
\newblock \showarticletitle{Image-based recommendations on styles and
  substitutes}. In \bibinfo{booktitle}{\emph{Proceedings of the 38th
  international ACM SIGIR conference on research and development in information
  retrieval}}. \bibinfo{pages}{43--52}.
\newblock


\bibitem[Ning and Karypis(2011)]%
        {ning2011slim}
\bibfield{author}{\bibinfo{person}{Xia Ning} {and} \bibinfo{person}{George
  Karypis}.} \bibinfo{year}{2011}\natexlab{}.
\newblock \showarticletitle{Slim: Sparse linear methods for top-n recommender
  systems}. In \bibinfo{booktitle}{\emph{2011 IEEE 11th International
  Conference on Data Mining}}. IEEE, \bibinfo{pages}{497--506}.
\newblock


\bibitem[Ning and Karypis(2012)]%
        {ning2012sparse}
\bibfield{author}{\bibinfo{person}{Xia Ning} {and} \bibinfo{person}{George
  Karypis}.} \bibinfo{year}{2012}\natexlab{}.
\newblock \showarticletitle{Sparse linear methods with side information for
  top-n recommendations}. In \bibinfo{booktitle}{\emph{Proceedings of the sixth
  ACM conference on Recommender systems}}. \bibinfo{pages}{155--162}.
\newblock


\bibitem[Pan et~al\mbox{.}(2019)]%
        {pan2019warm}
\bibfield{author}{\bibinfo{person}{Feiyang Pan}, \bibinfo{person}{Shuokai Li},
  \bibinfo{person}{Xiang Ao}, \bibinfo{person}{Pingzhong Tang}, {and}
  \bibinfo{person}{Qing He}.} \bibinfo{year}{2019}\natexlab{}.
\newblock \showarticletitle{Warm up cold-start advertisements: Improving ctr
  predictions via learning to learn id embeddings}. In
  \bibinfo{booktitle}{\emph{Proceedings of the 42nd International ACM SIGIR
  Conference on Research and Development in Information Retrieval}}.
  \bibinfo{pages}{695--704}.
\newblock


\bibitem[Popescul et~al\mbox{.}(2013)]%
        {popescul2013probabilistic}
\bibfield{author}{\bibinfo{person}{Alexandrin Popescul},
  \bibinfo{person}{Lyle~H Ungar}, \bibinfo{person}{David~M Pennock}, {and}
  \bibinfo{person}{Steve Lawrence}.} \bibinfo{year}{2013}\natexlab{}.
\newblock \showarticletitle{Probabilistic models for unified collaborative and
  content-based recommendation in sparse-data environments}.
\newblock \bibinfo{journal}{\emph{arXiv preprint arXiv:1301.2303}}
  (\bibinfo{year}{2013}).
\newblock


\bibitem[Qin et~al\mbox{.}(2021)]%
        {qin2021neural}
\bibfield{author}{\bibinfo{person}{Zhen Qin}, \bibinfo{person}{Le Yan},
  \bibinfo{person}{Honglei Zhuang}, \bibinfo{person}{Yi Tay},
  \bibinfo{person}{Rama~Kumar Pasumarthi}, \bibinfo{person}{Xuanhui Wang},
  \bibinfo{person}{Mike Bendersky}, {and} \bibinfo{person}{Marc Najork}.}
  \bibinfo{year}{2021}\natexlab{}.
\newblock \showarticletitle{Are Neural Rankers still Outperformed by Gradient
  Boosted Decision Trees?}
\newblock  (\bibinfo{year}{2021}).
\newblock


\bibitem[Reimers and Gurevych(2019)]%
        {reimers-2019-sentence-bert}
\bibfield{author}{\bibinfo{person}{Nils Reimers} {and} \bibinfo{person}{Iryna
  Gurevych}.} \bibinfo{year}{2019}\natexlab{}.
\newblock \showarticletitle{Sentence-BERT: Sentence Embeddings using Siamese
  BERT-Networks}. In \bibinfo{booktitle}{\emph{Proceedings of the 2019
  Conference on Empirical Methods in Natural Language Processing}}.
  \bibinfo{publisher}{Association for Computational Linguistics}.
\newblock
\urldef\tempurl%
\url{https://arxiv.org/abs/1908.10084}
\showURL{%
\tempurl}


\bibitem[Rendle(2010)]%
        {rendle2010factorization}
\bibfield{author}{\bibinfo{person}{Steffen Rendle}.}
  \bibinfo{year}{2010}\natexlab{}.
\newblock \showarticletitle{Factorization machines}. In
  \bibinfo{booktitle}{\emph{2010 IEEE International conference on data
  mining}}. IEEE, \bibinfo{pages}{995--1000}.
\newblock


\bibitem[Rendle et~al\mbox{.}(2020)]%
        {rendle2020neural}
\bibfield{author}{\bibinfo{person}{Steffen Rendle}, \bibinfo{person}{Walid
  Krichene}, \bibinfo{person}{Li Zhang}, {and} \bibinfo{person}{John
  Anderson}.} \bibinfo{year}{2020}\natexlab{}.
\newblock \showarticletitle{Neural collaborative filtering vs. matrix
  factorization revisited}. In \bibinfo{booktitle}{\emph{Fourteenth ACM
  Conference on Recommender Systems}}. \bibinfo{pages}{240--248}.
\newblock


\bibitem[Rendle et~al\mbox{.}(2022)]%
        {rendle2022revisiting}
\bibfield{author}{\bibinfo{person}{Steffen Rendle}, \bibinfo{person}{Walid
  Krichene}, \bibinfo{person}{Li Zhang}, {and} \bibinfo{person}{Yehuda Koren}.}
  \bibinfo{year}{2022}\natexlab{}.
\newblock \showarticletitle{Revisiting the performance of ials on item
  recommendation benchmarks}. In \bibinfo{booktitle}{\emph{Proceedings of the
  16th ACM Conference on Recommender Systems}}. \bibinfo{pages}{427--435}.
\newblock


\bibitem[Sharma et~al\mbox{.}(2015)]%
        {sharma2015feature}
\bibfield{author}{\bibinfo{person}{Mohit Sharma}, \bibinfo{person}{Jiayu Zhou},
  \bibinfo{person}{Junling Hu}, {and} \bibinfo{person}{George Karypis}.}
  \bibinfo{year}{2015}\natexlab{}.
\newblock \showarticletitle{Feature-based factorized bilinear similarity model
  for cold-start top-n item recommendation}. In
  \bibinfo{booktitle}{\emph{Proceedings of the 2015 SIAM International
  Conference on Data Mining}}. SIAM, \bibinfo{pages}{190--198}.
\newblock


\bibitem[Steck(2019)]%
        {steck2019embarrassingly}
\bibfield{author}{\bibinfo{person}{Harald Steck}.}
  \bibinfo{year}{2019}\natexlab{}.
\newblock \showarticletitle{Embarrassingly shallow autoencoders for sparse
  data}. In \bibinfo{booktitle}{\emph{The World Wide Web Conference}}.
  \bibinfo{pages}{3251--3257}.
\newblock


\bibitem[Steck(2020)]%
        {steck2020autoencoders}
\bibfield{author}{\bibinfo{person}{Harald Steck}.}
  \bibinfo{year}{2020}\natexlab{}.
\newblock \showarticletitle{Autoencoders that don't overfit towards the
  identity}.
\newblock \bibinfo{journal}{\emph{Advances in Neural Information Processing
  Systems}}  \bibinfo{volume}{33} (\bibinfo{year}{2020}),
  \bibinfo{pages}{19598--19608}.
\newblock


\bibitem[Steck et~al\mbox{.}(2020)]%
        {steck2020admm}
\bibfield{author}{\bibinfo{person}{Harald Steck}, \bibinfo{person}{Maria
  Dimakopoulou}, \bibinfo{person}{Nickolai Riabov}, {and} \bibinfo{person}{Tony
  Jebara}.} \bibinfo{year}{2020}\natexlab{}.
\newblock \showarticletitle{ADMM SLIM: Sparse Recommendations for Many Users}.
  In \bibinfo{booktitle}{\emph{Proceedings of the 13th International Conference
  on Web Search and Data Mining}}. \bibinfo{pages}{555--563}.
\newblock


\bibitem[Van{\v{c}}ura and Kord{\'\i}k(2021)]%
        {vanvcura2021deep}
\bibfield{author}{\bibinfo{person}{Vojt{\v{e}}ch Van{\v{c}}ura} {and}
  \bibinfo{person}{Pavel Kord{\'\i}k}.} \bibinfo{year}{2021}\natexlab{}.
\newblock \showarticletitle{Deep variational autoencoder with shallow parallel
  path for top-N recommendation (VASP)}. In
  \bibinfo{booktitle}{\emph{Artificial Neural Networks and Machine
  Learning--ICANN 2021: 30th International Conference on Artificial Neural
  Networks, Bratislava, Slovakia, September 14--17, 2021, Proceedings, Part V
  30}}. Springer, \bibinfo{pages}{138--149}.
\newblock


\bibitem[Vartak et~al\mbox{.}(2017)]%
        {vartak2017meta}
\bibfield{author}{\bibinfo{person}{Manasi Vartak}, \bibinfo{person}{Arvind
  Thiagarajan}, \bibinfo{person}{Conrado Miranda}, \bibinfo{person}{Jeshua
  Bratman}, {and} \bibinfo{person}{Hugo Larochelle}.}
  \bibinfo{year}{2017}\natexlab{}.
\newblock \showarticletitle{A meta-learning perspective on cold-start
  recommendations for items}.
\newblock \bibinfo{journal}{\emph{Advances in neural information processing
  systems}}  \bibinfo{volume}{30} (\bibinfo{year}{2017}).
\newblock


\bibitem[Volkovs et~al\mbox{.}(2017)]%
        {volkovs2017dropoutnet}
\bibfield{author}{\bibinfo{person}{Maksims Volkovs}, \bibinfo{person}{Guangwei
  Yu}, {and} \bibinfo{person}{Tomi Poutanen}.} \bibinfo{year}{2017}\natexlab{}.
\newblock \showarticletitle{Dropoutnet: Addressing cold start in recommender
  systems}.
\newblock \bibinfo{journal}{\emph{Advances in neural information processing
  systems}}  \bibinfo{volume}{30} (\bibinfo{year}{2017}).
\newblock


\bibitem[Wang and Isola(2020)]%
        {10.5555/3524938.3525859}
\bibfield{author}{\bibinfo{person}{Tongzhou Wang} {and}
  \bibinfo{person}{Phillip Isola}.} \bibinfo{year}{2020}\natexlab{}.
\newblock \showarticletitle{Understanding Contrastive Representation Learning
  through Alignment and Uniformity on the Hypersphere}. In
  \bibinfo{booktitle}{\emph{Proceedings of the 37th International Conference on
  Machine Learning}} \emph{(\bibinfo{series}{ICML'20})}.
  \bibinfo{publisher}{JMLR.org}, Article \bibinfo{articleno}{921},
  \bibinfo{numpages}{11}~pages.
\newblock


\bibitem[Wang et~al\mbox{.}(2023)]%
        {equal}
\bibfield{author}{\bibinfo{person}{Wenjie Wang}, \bibinfo{person}{Xinyu Lin},
  \bibinfo{person}{Liuhui Wang}, \bibinfo{person}{Fuli Feng},
  \bibinfo{person}{Yinwei Wei}, {and} \bibinfo{person}{Tat-Seng Chua}.}
  \bibinfo{year}{2023}\natexlab{}.
\newblock \showarticletitle{Equivariant Learning for Out-of-Distribution
  Cold-start Recommendation}. In \bibinfo{booktitle}{\emph{Proceedings of the
  31st ACM International Conference on Multimedia}}. \bibinfo{pages}{903--914}.
\newblock


\bibitem[Wei et~al\mbox{.}(2021)]%
        {CLCRec}
\bibfield{author}{\bibinfo{person}{Yinwei Wei}, \bibinfo{person}{Xiang Wang},
  \bibinfo{person}{Li Qi}, \bibinfo{person}{Liqiang Nie}, \bibinfo{person}{Yan
  Li}, \bibinfo{person}{Xuanqing Li}, {and} \bibinfo{person}{Tat-Seng Chua}.}
  \bibinfo{year}{2021}\natexlab{}.
\newblock \showarticletitle{Contrastive Learning for Cold-start
  Recommendation}. In \bibinfo{booktitle}{\emph{Proceedings of the 29th ACM
  International Conference on Multimedia}}.
\newblock


\bibitem[Xu et~al\mbox{.}(2022)]%
        {xu2022alleviating}
\bibfield{author}{\bibinfo{person}{Xiaoxiao Xu}, \bibinfo{person}{Chen Yang},
  \bibinfo{person}{Qian Yu}, \bibinfo{person}{Zhiwei Fang},
  \bibinfo{person}{Jiaxing Wang}, \bibinfo{person}{Chaosheng Fan},
  \bibinfo{person}{Yang He}, \bibinfo{person}{Changping Peng},
  \bibinfo{person}{Zhangang Lin}, {and} \bibinfo{person}{Jingping Shao}.}
  \bibinfo{year}{2022}\natexlab{}.
\newblock \showarticletitle{Alleviating Cold-start Problem in CTR Prediction
  with A Variational Embedding Learning Framework}. In
  \bibinfo{booktitle}{\emph{Proceedings of the ACM Web Conference 2022}}.
  \bibinfo{pages}{27--35}.
\newblock


\bibitem[Zhang et~al\mbox{.}(2018)]%
        {zhang2018discrete}
\bibfield{author}{\bibinfo{person}{Yan Zhang}, \bibinfo{person}{Hongzhi Yin},
  \bibinfo{person}{Zi Huang}, \bibinfo{person}{Xingzhong Du},
  \bibinfo{person}{Guowu Yang}, {and} \bibinfo{person}{Defu Lian}.}
  \bibinfo{year}{2018}\natexlab{}.
\newblock \showarticletitle{Discrete deep learning for fast content-aware
  recommendation}. In \bibinfo{booktitle}{\emph{Proceedings of the eleventh ACM
  international conference on web search and data mining}}.
  \bibinfo{pages}{717--726}.
\newblock


\bibitem[Zhao et~al\mbox{.}(2022)]%
        {zhao2022improving}
\bibfield{author}{\bibinfo{person}{Xu Zhao}, \bibinfo{person}{Yi Ren},
  \bibinfo{person}{Ying Du}, \bibinfo{person}{Shenzheng Zhang}, {and}
  \bibinfo{person}{Nian Wang}.} \bibinfo{year}{2022}\natexlab{}.
\newblock \showarticletitle{Improving item cold-start recommendation via
  model-agnostic conditional variational autoencoder}. In
  \bibinfo{booktitle}{\emph{Proceedings of the 45th International ACM SIGIR
  Conference on Research and Development in Information Retrieval}}.
  \bibinfo{pages}{2595--2600}.
\newblock


\bibitem[Zhu et~al\mbox{.}(2021)]%
        {10.1145/3437963.3441820}
\bibfield{author}{\bibinfo{person}{Ziwei Zhu}, \bibinfo{person}{Yun He},
  \bibinfo{person}{Xing Zhao}, \bibinfo{person}{Yin Zhang},
  \bibinfo{person}{Jianling Wang}, {and} \bibinfo{person}{James Caverlee}.}
  \bibinfo{year}{2021}\natexlab{}.
\newblock \showarticletitle{Popularity-Opportunity Bias in Collaborative
  Filtering}. In \bibinfo{booktitle}{\emph{Proceedings of the 14th ACM
  International Conference on Web Search and Data Mining}} (Virtual Event,
  Israel) \emph{(\bibinfo{series}{WSDM '21})}. \bibinfo{publisher}{Association
  for Computing Machinery}, \bibinfo{address}{New York, NY, USA},
  \bibinfo{pages}{85–93}.
\newblock
\showISBNx{9781450382977}
\urldef\tempurl%
\url{https://doi.org/10.1145/3437963.3441820}
\showDOI{\tempurl}


\end{thebibliography}

\end{document}